\documentclass[aps, reprint, nofootinbib]{revtex4-1}
\usepackage{amsmath, latexsym, amssymb, hyperref, graphicx, color, slashed}
\definecolor{nicered}{rgb}{0.7,0.1,0.1}
\definecolor{nicegreen}{rgb}{0.1,0.5,0.1}
\hypersetup{colorlinks, citecolor=nicegreen,linkcolor=nicered}
\begin{document}

\newcommand{\Sec}[1]{ \medskip \noindent {\sl \bfseries #1}}
\newcommand{\subsec}[1]{ \medskip \noindent {\sl \bfseries #1}}
\newcommand{\Par}[1]{ \medskip \noindent {\em #1}}

\title{Particle physics: a personal view $^{\S}$}

\author{Goran Senjanovi\'c  }

\affiliation{Abdus Salam International Centre for Theoretical Physics, Trieste, Italy}

\affiliation{Arnold Sommerfeld Center, Ludwig-Maximilians University, Munich, Germany}

\begin{abstract}

  The reader surely knows what particles physics is about: finding building blocks of nature that appear elementary at a given time and study their interactions - so why in the world this essay? The problem is how to arrive at a fundamental theory, and maybe even more important, what is the theory supposed to do? I follow here a simple but profound prescription of Feynman for a true, self-contained theory, and then show how the Standard Model beautifully illustrates this prescription. That the issue is not trivial, is clear from the fact  that we had theories dominate our field for decades, in spite of completely failing Feynman's requirements and offering no clear predictions whatsoever. I next critically review two unique candidates for a Beyond Standard Model theory: the Left-Right Symmetric theory of electro-weak interactions and the minimal grand unified theory based on SU(5) gauge symmetry. I also comment on some generic aspects of grand unification, and what hope we may have in testing it. 
\end{abstract}


\maketitle

{
 \renewcommand{\thefootnote}%
   {\fnsymbol{footnote}}
 \footnotetext[4]{Based on the closing talk of the International conference 'LHC Days in Split 2024, Hvar, Croatia}
}

\section{Prologue} \label{Prologue}

      I was asked by the organisers of the LHC Days in Split 2024 conference to give a closing talk that should summarise the essence of what particle physics is all about, and offer some insight of what future may bring - a tall task.  It is hard to have that insight no matter how much one reflects, since one is necessarily biased by one's taste and a sense of direction - there are no clear experimental hints on what the new theory may be about.  Thus I decided in this essay to look for inspiration by recalling the development of  the Standard Model (SM), a perfect example of a simple but profound theory based on fundamental principles of gauge symmetry and its spontaneous breaking through the Higgs mechanism. 
      
      I recall  how from these principles one even ended up with a theory of the origin of mass of all present day elementary particles.  Well, almost all of them. While the SM beautifully accounts for the masses of W and Z bosons, the Higgs boson mass and the masses of charged fermions, through the Higgs-Weinberg mechanism,  it fails, however, when it comes to neutrino - it predicts it massless. This is the only shortcoming of the SM, its only incompleteness, and as such offers a unique window into the new physics. 
      
      Let us be clear here. What I mean is that the SM is a theory of strong and electro-weak interactions, not an imaginary theory of all phenomena. Issues like say the origin of dark matter or genesis are simply outside its scope. On the other hand, weak interaction is all about neutrino and thus the SM wrong prediction of vanishing neutrino mass is its true shortcoming and as such arguably our best door into the physics Beyond the Standard Model (BSM). \\
      
      Before going on: For a quick recent review of the history of the development of the SM, I strongly recommend a nice essay by one of its fathers~\cite{Glashow:2022uat}), while a reader in need of a pedagogical treatise of ideas offered here could profit from~\cite{Melfo:2021wry} (a short {\it expose} can be found in~\cite{Senjanovic:2020rcq}).\\
      
      One often claims that there are deep problems in the SM, such as the smallness of strong CP violations or the smallness of the weak scale and of the vacuum energy.  I argue below that these are not problems whatsoever and that the SM is a perfectly consistent theory of these issues. They do remain interesting questions, although one should be careful when pretending to have a theory of the values of the parameters of a theory - there is a serious danger that one ends up like Kepler computing the distances of the planets from the Sun.  After all, the first and foremost task of physics as a natural philosophy, as Newton coined it nicely, is to correlate apparently different physical phenomena, and not to worry why the numbers are what they are, or why they are smaller than what we would like them to be. 
      
      The obsession with the latter issue, usually defined as the issue of naturalness, has repeatedly led to claims that the new physics had to lie already at LEP energies, not to speak of the LHC ones. In particular, the low energy supersymmetry, which would somehow magically tell us why Higgs is light, was already supposed to be seen. This claim has now led to a backlash and an equally unphysical claim that the theory is dead. In other words, based on not seeing any of the supersymmetric partners of SM particles, it was first argued that supersymmetry had to lie below TeV energies, to turn into a proclamation of the theory being ruled out - whereas all that happened is that the limits on its scale are being raised. \\
      
      Let me emphasise that I too had fallen in love with low energy supersymmetry more than 40 yeas ago and that I have vested interest in it.  Bill Marciano and I had even argued in favour of a heavy top quark, with the mass on the order of 200 GeV~\cite{Marciano:1981un,Senjanovic:2012zt}, when almost everybody was dead sure that its mass would have lied below some 20 TeV or so. This was back in 1981 when the weak mixing angle was wrongly extrapolated from low energy, while low energy supersymmetry unification required it larger~\cite{Dimopoulos:1981yj}, and the heavy top would have boosted its value . This had provided another argument in favour of low energy supersymmetry, besides its role in hopefully keeping Higgs light. Grand unification was at its peak in those days, and 'everybody rushed underground' to look for proton decay, as Morris Goldhaber had put it nicely. - and so, unification of gauge couplings was on the minds of most people. 
      
      Some ten years later, LEP would confirm the larger value of the weak mixing angle, prophetically predicted by low energy supersymmetry which turned into a dogma - and a few years later the top quark mass would fit nicely with what we preached. The trouble is that the explicit realisation, the minimal supersymmetric standard model ended up having more than 100 parameters, and as a phenomenological theory it failed completely - and yet its failure led almost to its cult. Talk about the politics of science. \\

       The strangest aspect of all of this is that over the years the success of the SM became spectacular and it turned into a high precision theory - and yet, almost everyone kept insisting that new physics had to lie at low energies. If anything, the SM message is the opposite one, it would be a miracle to have new phenomena occur at the LHC. We all hope for that miracle, needless to say, but nonetheless it would be a miracle. Fortunately, there is a simple phenomenological low energy phenomenon that would cry for new physics, if not at the LHC, then at the next hadron collider - and that is the observation of the neutrinoless double beta decay in near future and the chirality of outgoing electrons be Right-Handed (RH). 
       
       This is one of the central messages of what follows below. In that case, one could expect a plethora of physical processes at the hadron colliders. I hope that this {\it per se} motivates the reader to keep reading this little review of the basics of the SM and BSM physics, although my scope is more ambitious - I wish to address an issue of what it really means to construct a physical theory and how one goes about it, with the help of Feynman and his genius of teaching us what science is all about. \\

 \section{Making of a theory} \label{theory}
 
   Feynman, in his usual simple but deep manner, explains the construction of theories of nature~\cite{feynman}, which I summarise somewhat freely:
   
 \begin{itemize}
\item One makes a guess - call it a principle, hypothesis, idea, whatever. An example: equivalence principle that Einstein liked. 
\item  You then construct the minimal formulation based on that guess. 
\item  Then one computes the predictions. If they are unambiguous, we have a self-contained, predictive theory.
\item  Then comes experiment. If this good theory manages not to be falsified for a long time, the rest will be history.  Needless to say, the consequences of the theory need to be clear and precise. 
\end{itemize}

Or if you wish, these are his own words: {\it “OK, now that’s the present situation. Now I’m going to discuss how we would look for a new law. In general, we look for a new law by the following process. First, we guess it.

Then, we compute– well, don’t laugh, that’s really true. Then we compute the consequences of the guess, to see what, if this is right, if this law that we guessed is right, we see what it would imply. And then we compare those computation results to nature. Or we say, compare to experiment or experience. Compare it directly with observation, to see if it works.

If it disagrees with experiment, it’s wrong. And that simple statement is the key to science. It doesn’t make any difference how beautiful your guess is, it doesn’t make any difference how smart you are, who made the guess, or what his name is. If it disagrees with experiment, it’s wrong. That’s all there is to it.”}\\

The crucial thing is the clarity and predictivity, as he puts it nicely: {“\it I must also point out to you that you cannot prove a vague theory wrong. If the guess that you make is poorly expressed and rather vague, and the method that you used for figuring out the consequences is rather vague, you’re not sure, and you just say I think everything is because it’s all due to moogles, and moogles do this and that, more or less. So I can sort of explain how this works. Then you say that that theory is good, because it can’t be proved wrong.

If the process of computing the consequences is indefinite, then with a little skill, any experimental result can be made to look like an expected consequence. ”}\\

A nice example of a vague theory that cannot be tested: the string theory. It never made any unambiguous prediction and yet it had been coined as the Final theory (sounds familiar?) and The theory of everything - everything, imagine. Its problem is precisely its vagueness that bothers Feynman - being defined in ten space-time dimensions, it is a theory with basically infinitely many vacua (see more below). This had taken the field to the road of faith to either guess the right vacuum by divine inspiration or use its enormous landscape to argue that we should not even look for a fundamental theory in our four-dimensional world - after all this theory of everything was supposed to take us to a promised land. 

What is really strange, is that it emerged as the final theory some forty years ago, at the time when Quantum Field Theory had not yet reached its peak. W boson discovery was still quite fresh and we would have to wait almost 40 years for the Higgs boson discovery.  Think about it, you were supposed to give up what worked beautifully, what led to clear predictions, what you understood deeply for something ambiguous, ill understood, murky. 

In any case, it is amusing, to put it mildly, that the theory of nothing was proclaimed the theory of everything.
Moreover, by that time there were already indications that neutrinos were massive and thus that the SM was incomplete, opening a window into a new physics, likely based on QFT. Talk about populism of science.\\

A nice example of a clear, self-contained theory: the Standard Model of strong and electroweak interactions, with the guess being the gauge principle and the Higgs mechanism. It is worth recapturing briefly the history of this theory, to get a glimpse into its emergence from the confusion of the dark ages of bootstrap and Regge poles - the era of the '60s when both the Quantum Field Theory and fundamental physics were proclaimed dead. It is probably hard to imagine for a young reader, whose education is all about QFT, something so silly - but the example of string theory some 20 years later shows again the danger of fashions that tend to plague high energy physics when one lacks hints from experiment. \\
 
 \section{Standard Model: the essence } \label{SM}

  In what follows I will speak only about the electroweak part of the SM, to be precise. In a nutshell, the SM is based on a fundamental theoretical principle and a profound fact of nature
 
 \begin{itemize}
 
 \item Gauge principle and Higgs mechanism 
 
 \item Maximal parity violation in charged currents.
 
 \end{itemize}
 
 While we always emphasise the former feature,  in my experience the latter one, the fact that only left-handed fermions participate in charged weak interaction, is not appreciated enough, neither by teachers of the SM nor by  its students - and yet it is absolutely crucial. Not for the structure of the SM of course, for you could equally well build the $SU(2) \times U(1)$ gauge theory in parity conserving world, simply by taking both LH and RH fermions to be weak doublets - that part is trivial. The problem, though, would be spontaneous symmetry breaking, the Higgs mechanism aspect - it simply would not work, not in a sense of being predictive. 
 
 To understand better the point I wish to make, let us first recapitulate the essential features of the SM, those that we all know and love. Even better, let us go back to pre SM attempt at the gauge theory of electroweak interaction, the $SU(2)$ model of Glashow. In this case the electromagnetic charge is both predicted and automatically quantised
  \begin{equation}\label{su2charge}
 Q_{em} = T_3\,,
 \end{equation}
 where $T_3$ stand for the neutral $SU(2)$ generator, and thus its eigenvalues are integer multiples of the minimal charge 
  \begin{equation}\label{su2charge}
 q_{em} = \pm \frac{1}{2}\,.
 \end{equation}
 A miracle of charge quantisation in nature would be automatically explained by the minimal gauge theory of both electromagnetic and weak phenomena, albeit wrongly. An example of a great tragedy of science, beautiful theories killed by ugly facts of nature? Maybe, but then maybe not - and we will come to this below. The way out of this impasse is remarkably simple, a kind of Nike approach: just do it. You all know it, Glashow added a $U(1)$ factor~\cite{Glashow:1961tr}, and the rest became history when Weinberg~\cite{Weinberg:1967tq} and Salam~\cite{Salam:1968rm} used the Higgs mechanism to turn this model into a complete, predictive, self-contained theory.  Unfortunately, while charges can be accounted for, they cannot be predicted and sadly they are not quantised since 
  \begin{equation}\label{su2charge}
 Q_{em} = T_3 + \frac{Y}{2}\,,
 \end{equation}
 where $Y$ stand for the $U(1)$ generator, with arbitrary values. \\
 
 The fermion structure is completely LR asymmetric 
  \begin{equation}\label{smferm}
  \left(\begin{array}{c} u \\ d \end{array} \right)_L, \,\,\, \left(\begin{array}{c} \nu \\ e \end{array} \right)_L;\,\,\,\,\, u_R,\, \,d_R,\,\, e_R\,.
 \end{equation}
 Notice the absence of the RH neutrino due to minimality principle. Keep in mind that the minimality, besides the gauge principle itself, is essential for the sake of predictivity. \\

\subsection{Predictions from gauge principle} \label{gaugepredictions}

The gauge principle leads to two important predictions

 \begin{itemize}
 
 \item universal neutral current
   \begin{equation}\label{zcurrent}
J_{Z}^\mu = \bar f (T_3 - \sin^2\theta_W \,Q_{em}) \gamma^\mu f,
\end{equation}
 which follows from the assumption that there is a massless gauge boson, the photon $A$, corresponding to the em current 
 $J_{A}^\mu = \bar f Q_{em} \gamma^\mu f$.
  
  \item determination of W-boson mass
  \begin{equation}\label{wmass}
  \frac {G_F}{\sqrt 2} = \frac { \pi \alpha}{2 M_W^2 \sin^2\theta_W},
 \end{equation}
 which gives $M_W \sin\theta_W \simeq 40 \,GeV$. This would eventually lead to a celebrated  predicton $M_W  \simeq 80\, GeV$ once the weak mixing angle got  experimentally determined  to be $ \sin\theta_W \simeq 1/2$.
 \end{itemize}

These are truly remarkable predictions, from a simple fundamental theory - and yet nobody cared. I  have always wondered if I would have recognised its importance had I been around. What is really amusing is that Glashow himself does not cite his masterpiece, that started it all, in his 1964 paper with Bjorken~\cite{Bjorken:1964gz} suggesting the existence of the charm quark from quark-lepton symmetry.   It is interesting that Feynman in his recipe how to come with a fundamental theory offers no wisdom on what to do if you get completely ignored. I often wonder why he too ignored Glashow's great work. \\

\subsection{Predictions from Higgs mechanism} \label{higgspredictions}

The problem that Glashow had was how to account for W and Z masses since explicit breaking makes the theory non-renormalisable. This was solved by Weinberg in 1967~\cite{Weinberg:1967tq} through the Higgs mechanism, although he could deal with leptons only - the quark picture was completed by Glashow (again), Illiopoulos and Maiani in 1969~\cite{Glashow:1970gm}, who showed that charm quark was necessary, not just appealing. What emerged was extraordinary: a single Higgs doublet suffices to produce 
all SM particle masses which led to a number of predictions

 \begin{itemize}
 
 \item massless photon and conserved em current 
 
 \item W-Z mass ratio determined $M_W = M_Z \cos\theta_W$, predicting $ M_Z \simeq 90\, GeV$
 
 \item Higgs boson decay rates determined by particle masses, e.g. decays into charged fermions
   \begin{equation}\label{higgstoffbar}
\Gamma (h \to f \bar f) \propto \left(\frac{m_f}{M_W}\right)^2 m_h
\end{equation}
  \item Higgs boson mass related to scattering amplitudes, just like W mass
   \begin{equation}\label{higgsscattering}
A (h +h \to h + h) \propto \left(\frac{m_h}{M_W}\right)^2 
\end{equation}

\item neutrinos are massless.
 
 \end{itemize}

This is arguably one of the greatest pages in the history of modern physics: a fundamental theory of electroweak interactions turned also into a theory of the origin of mass. Today we know that W, Z boson and third generation of charged fermions get their masses from the Higgs mechanism, and it is a safe bet (I am willing to put my money where my mouth is) to assume the same for the first two generations.\\  

\paragraph{The role of parity violation} The reason this worked is the maximal violation of parity. To appreciate this fully, imagine the left-right symmetric world, which would have forced Glashow to assume both LH and RH fermions to be doublets - making Weinberg's task basically impossible. The Higgs doublet would be useless for fermion masses, which would have surely discouraged him to use it. But even if would have used it, the problem of realistic fermion masses would be formidable. 

First of all, fermions get direct mass terms, equal for both up and down components - not only would up and down quarks have same masses, but also neutrino and electron. In order to break this symmetry, a neutral Higgs triplet would be needed, and then one would have to arrange for the (almost total) cancellation of neutrino masses. In turn, W and Z masses would not be correlated, and Yukawa couplings would not be directly related to fermion masses - all predictions would be gone. Worse, there would be problems with strangeness conservation in neutral currents - all hell would break loose. In short, the reason electroweak unification works as a renormalisable theory is maximal parity violation in charged weak interaction.\\

\paragraph{Neutrino mass} The SM structure leads to one of its clearest predictions, massless neutrino. Of course, the crucial reason is the minimality of the theory, which imposes the absence of the RH neutrino, never seen in nature - and then in turn, the gauge symmetry also forbids a possible Majorana mass term for neutrino. In the imaginary LR symmetric world, though, neutrinos would be massive, an important fact worth keeping in mind. \\

\subsection{SM problems?} \label{problems}

If there is one problem with the SM, it is sadly a total lack of problems. The SM is an extraordinary high precision theory, working too well, a discouraging fact for a searcher of the theory beyond. So why does everyone speak of the SM problems? To convince the reader, lets us go through these so-called problems, the ones of naturalness: why are numbers what they are?\\

\paragraph{Hierarchy problem?}

Unlike fermion masses which are protected in perturbation theory, in a sense that the loop contributions are necessarily proportional to the tree level value, the scalar mass terms, such as say the Higgs boson mass, are sensitive to large scales due to apparent quadratic divergences. If we were not allowed, or better obliged, to renormalise tree level parameter values, this could be a problem, but it is obviously just a dislike for fine-tuning, an emotional or aesthetic argument, not a physical one.  In order to sound profound, one then often asks a philosophical question of why is Higgs mass scale, or better the weak interaction scale, is so much smaller than the Planck or grand unified (GUT) scale. This is a dangerous question for it is not even clear how to answer it - after all. we know everything about the weak scale and basically nothing about Planck or GUT scales. \\

To appreciate better the problem of claiming that the smallness of Higgs mass is a problem, it is enough to think about the absurdity of claiming that explanation lies in low energy supersymmetry. Think about it - if the Higgs lightness is a problem, so is equally the smallness of supersymmetry breaking scale. In other words, low energy supersymmetry only aggravates the issues, by basically doubling the problem, for now we must understand why the new scale is similarly so much smaller than the Planck and GUT scales. The only thing low energy supersymmetry does is to allow fine-tuning of the Higgs mass at the tree level once for all - the loops are then automatically small -  but nonetheless it must be fine-tuned. \\

This is made clear in the context of grand unification where the fine-tuning becomes more manifest since one has to literally cancel enormous contributions to the Higgs mass term, contributions proportional to the GUT scale, some 13 orders of magnitude (or more) bigger than the weak scale - and one has to do it order by order in perturbation theory. If the theory is made supersymmetric, and moreover all the way to weak scale energies, the tree level fine-tuning suffices. This is surely appealing but it comes at a great price as I said, for now we face another miracle of smallness of scales. As I mentioned in the Prologue, in the early eighties low energy GUT picture led to an exciting prediction of the correct value of the weak mixing angle, and even tied it to the largeness of the top quark mass - which led to the low energy supersymmetry craze and even promises of finding it first at LEP and then LHC energies. This is often called naturalness criterion: increasing the supersymmetry breaking scale starts requiring minor fine tuning at loop level.  In spite of my great interest in low energy supersymmetry, and even the love for it, I for one was deeply unhappy when it became a dogma - for decades it was not consider exotics at the LHC in spite of no shred of evidence in its favour - but I am now equally unhappy with its necrologies. 

With every day that passes, the SM Higgs sector looks more and more solidly established. It arguably makes no sense to claim that new physics must be around the corner - its using the success of the theory against it.\\

\paragraph{Cosmological constant problem?}

This is very similar to the hierarchy problem. The cosmological constant, which is the vacuum energy, is many orders of magnitude smaller than any particle physics scale. Again, it is not a problem since it is not calculable, and so I will not dwell on it any more. I should only mention that Dvali has been arguing that it must be strictly zero from the requirement of unitarity of quantum gravity~\cite{Dvali:2020etd}.  Although this makes zero a special number, ironically it makes the question of its smallness even more pronounced - after all,  zero is the smallest small number. \\

\paragraph{Strong CP problem?}

This is yet another not existing problem, I claim. It has to do with the strong CP parameter being rather tiny, again argued not to be natural. It is, however, rather different from the Higgs mass question, and it requires a careful analysis of what naturalness means, so let me address it here. What follows has been borrowed heavily from my essay~\cite{Senjanovic:2020pnq}.

Imagine a physical parameter,   
    $\theta$ divided in tree-level and loop corrections 
  \begin{equation}\label{theta}
\theta = \theta_{tree} + \theta_{loop}.
 \end{equation}
 In case 
  \begin{equation}\label{loop}
 \theta_{loop} \gg \theta_{exp},.
 \end{equation}
where $\theta_{exp}$ is an experimental value or just a bound, we say that the parameter 
$\theta$ is unnaturally small. It then requires the fine-tuned cancellation between 
$\theta_{tree}$ and $\theta_{loop}$, which, although admittedly ugly, is perfectly allowed both physically and mathematically. This has come to be known a naturalness, or a fine-tuning problem. When 
$\theta$ is dimension-full it is called a hierarchy problem, since it requires a hierarchy of scales - the situation with the Higgs boson mass discussed above.\\

 Although this is not a problem, since  it can always be done at the needed precision and at a chosen order of perturbation theory, it is certainly a troubling issue. 
There is no issue of unnaturalness however when a protective symmetry keeps $\theta_{loop}$ small in perturbation theory, such as say chiral symmetry in the limit of vanishing fermion mass. This is called technical naturalness~\cite{tHooft:1979rat}, and no fine tuning is needed in such a case. Besides technical naturalness, though, there is a more natural, intuitive  definition of being natural - something that happens in our daily experience.  Small fermion masses are technically
 natural, sure, but small fermion mass is truly natural, since most of them are small. \\
 
 Let us then apply this reasoning to strong CP violation. It is known that the effective strong CP parameter $\bar \theta$ is extremely small, 
   $\bar \theta_{exp} \lesssim 10^{-10}$, which prompted the claim of the strong CP problem, by proclaiming that naturally this parameter should be much larger, even of order one (this is silly of course, numbers of order one are rather rare in the SM). While this may be debatable, there is no problem in a technical sense of large perturbative contributions to $\bar \theta$ - on the contrary, they are actually vanishingly small $\bar \theta_{loop} \lesssim 10^{-19}$~\cite{Ellis:1978hq}. To be precise, in principle $\bar \theta$ diverges but so mildly that the Planck scale cutoff implies this roughly tiny number~\cite{Ellis:1978hq}. In any case, there is no problem whatsoever with strong CP violation - a SM practitioner can happily work with $\bar \theta$ on top of usual weak CP violation, and she is guaranteed that nothing can go wrong. Of course, experiment can require going beyond this simple picture, but theoretically and conceptually the theory is perfectly OK. \\
   
   This said, it would surely nice to have a predictive theory of $\bar \theta$ parameter, although it is hard to understand how that would work in a perturbative BSM theory. One often claims that the Peccei-Quinn mechanism based on $U(1)_{PQ}$ global symmetry~\cite{Peccei:1977hh} and its associated pseudo-Goldstone boson, the axion~\cite{Weinberg:1977ma,Wilczek:1977pj}, predicts $\bar \theta = 0$, but that is simply wrong - the vanishing of $\bar \theta$ is simply traded for the {\it ad-hoc} exactness of $U(1)_{PQ}$ symmetry in the absence of instanton effects. Instead, the beauty of Peccei-Quinn mechanism lies in promoting a static variable $\bar \theta$ into a dynamic quantity, the vacuum expectation value of the axion field. This is commendable {\it per se}, but it does not tell us tell us why $\bar \theta$ is small. It does protect it in perturbation theory, due to symmetry arguments, but in the SM  the  smallness of $\bar \theta$ is structurally protected  in perturbation theory, as we have seen. As much as I like the axion picture, I feel equally uneasy about invoking a global symmetry, which is not even a true symmetry due to instanton breaking effects. It would be much nicer to have this symmetry be an accidental consequence of some deeper symmetry, such as grand unified gauge symmetry~\cite{Georgi:1981pu} or a family symmetry~\cite{Wilczek:1982rv,Chang:1984ip,Chang:1987hz}, but no simple and convincing model has emerged to this day. \\

\paragraph{Fermion mass problem?}

One often hears people talk even of the fermion mass problem, which simply makes no sense - the SM is a beautiful theory of the origin of particle masses. True, we do not predict values of the masses, or better mass ratios, but with the Higgs mechanism mass becomes a dynamical parameter - to every mass there is an associated well defined process, which allows us to probe the origin of mass. To ask that the theory determines its own parameters is almost an end of physics, for the task of science is to relate apparently distinct phenomena into a common framework, period not asking why numbers are what they are. \\

While it may be appealing to search for a BSM theory that can relate masses to each other, or masses to weak mixing angles~\cite{Weinberg:1977hb}, no true progress have ever been achieved. This said, I should admit that I too worked on this, both from symmetry principles~\cite{Mohapatra:1977rj} and the physics of extra dimensions~\cite{Rusjan:1989vm}. Looking back, I feel that it was futile, if for nothing else but for the fact that the patterns of masses and mixings in the SM are so chaotic, that it is hard to imagine, if possible at all, to make predictions based on simple and elegant principles. There is something else to keep in mind: in the SM the question itself makes no physical sense, since the only physical aspect of the theory sensitive to flavour mixing, the Cabibbo-Kobayashi-Maskawa mixing matrix does not depend on a particular form of up and down quark mass matrices. The issue becomes relevant un BSM theories where the currents associated with new gauge bosons do depend on individual quark mixing matrices. \\

\paragraph{Why three generations?}

Although maybe less often, nonetheless one tends to claim that the number of generations is also an issue to be understood. It is not clear how to do that, or even how to precisely pose the question - after all, one number is as good as another. Moreover, three generations in the SM are mandatory, as we all know, for the sake of CP violation, and were predicted by Kobayashi and Maskawa~\cite{Kobayashi:1973fv}. To me, this should suffice, for CP violation is a beautiful phenomenon and is essential for baryogenesis to work~\cite{Sakharov:1967dj}. True, in the case of leptogenesis~\cite{Fukugita:1986hr}, two generations could be sufficient due to extra complex phases in the Majorana picture, but then we could ask the question: why two generations? Still, even this is asked from the theory, to predict the number of generations, and here one often claims that a family symmetry, such as $SU(3)$ does the job. I don't have to explain how wrong this argument is, being simply a confusion between a cause and a consequence - if there were 2 generations, we would be motivated to assume $SU(2)$ symmetry, and so on. \\

There could be in principle a theoretical argument in favour of the uniqueness of number of generations being there. I personally got very excited when in the eighties it appeared that the natural way of compactifying extra six dimensions in a 10-dimensional space-time picture of string theory was based on Calabi-Yau spaces~\cite{Candelas:1985en}, and it seemed that there was a unique way of building such manifolds as algebraic varieties on complex manifolds. The number of generations in this picture is a topological invariant and it turned out that of some 20,000 such topologically distinct manifolds, only one supported three generations~\cite{Candelas:1987kf}. This was a remarkable result, but it was soon learned that the number of six dimensional Calabi-Yau spaces is actually infinite, and that there are other ways of compactifying, implying - at least to me - that the whole program was doomed. It is amusing how this failure, together with the failure to make the string theory a predictive renormalisable theory of gravity - in other words, a failure to make the string theory a theory of something - led to the claims that it is the Theory of Everything. I must admit here that when I started my journey to become a physicist I could have never, for the life of me, imagined that even physics can turn into a populist movement. Live and learn. \\

\subsection{Neutrino mass question}

For some fifty years or so, it has been preached that the above questions are problems of the SM and, as such, doors to physics beyond. I find this really strange, especially in view of the fact that the theory is incomplete, fortunately - it wrongly predicts vanishing of neutrino mass. This prediction lies at the core of the SM success of the Higgs mechanism and to me it is arguably the (The?) door to BSM.  Sheer logic and physical reasoning tells us to attack the theory where it fails phenomenologically - or is at least incomplete - rather than worrying about ill defined issues of aesthetic or philosophical nature. As I have argued in ~\cite{Senjanovic:2020pnq}, physics is supposed to be natural philosophy, not a philosophy of naturalness. \\

Of course, neutrino masses are rather tiny and thus new physics behind them could lie at astronomically high energies - strictly speaking, we have no argument to have BSM physics around the corner. That however is expected from the high precision success of the SM, and no desire of naturalness or such can change that fact - this is why I call neutrino mass an issue and not a problem, in spite being the only shortcoming of the SM. \\

  In other words, there is no theoretical, physical or experimental reason to have new physics reachable at near future hadron colliders, all claims on the contrary are just fake news.  Fortunately, there is a low energy process that could force the scale of new physics to be accessible at the LHC or the next hadron collider, i.e. neutrinoless double beta decay. If it were to be observed in near future and if the electrons in the final states were to be RH, the scale of new physics would have to lie at most at about 10 TeV or so - but more about it below.  I should stress that even if the electrons were to be LH, that would still not guarantee that this process is due to neutrino Majorana mass - it could still happen that it is due to some new high energy physics~\cite{Dvali:2023snt}.\\

\subsection{Road to BSM theory}\label{road}

Since new physics could be far from today's energies, what route should one take in our search for BSM? As I argued neutrino mass is a logical starting point, but even then, how to go about it? I would say that our best bet is to follow Feynman and look for fundamental theories with clear structural predictions of new physics. Arguably, the most natural candidates are 

 \begin{itemize}
 
 \item Left right symmetric theory 
 
 \item Grand unification 
 
 \end{itemize}

The reason I chose these two examples is simple: they satisfy nicely the criteria for a self-contained theory of nature, especially the LR model. They are both natural extensions of the SM, the former because it touches into the core of the breaking of parity in weak interaction, and the latter since it follows the original road that Glashow took, the road of unification - and both are based on fundamental principles of gauge symmetry and its spontaneous breakdown, the principles that led to it all. Both made structural predictions from the outset, and were not constructed to account for experimental facts {\it a posteriori}. Moreover, in spite of the theories  being old, there has been a significant progress in recent years in both of them. There is a profound difference though: The minimal left-right symmetric theory turns out to be highly predictive, while the minimal models of grand unifications fail and their realistic extensions lack predictivity to this day. \\

 \vskip 2 cm
 
   \section{LR symmetric theory}  \label{lr}  
   
   Left-right symmetric theory~\cite{Pati:1974yy,Mohapatra:1974gc,Senjanovic:1975rk,Senjanovic:1978ev}  emerged in the '70s as an attempt to attribute parity violation in nature to spontaneous breaking of left-right symmetry. It requires the minimal gauge group to be $SU(2)_L \times SU(2)_R \times U(1)_{B-L}$, augmented with left-right symmetry in terms of parity (P), but it can be as well substituted by charge conjugation (C). Its minimal fermion content is in turn determined to be left-right symmetric, completely opposite from the SM one
  \begin{equation}\label{lrferm}
  \left(\begin{array}{c} u \\ d \end{array} \right)_L, \,\,\, \left(\begin{array}{c} \nu \\ e \end{array} \right)_L;\,\,\,\,\, \left(\begin{array}{c} u \\ d \end{array} \right)_R, \,\,\, \left(\begin{array}{c} \nu \\ e \end{array} \right)_R\,.
 \end{equation}
   Two important predictions emerge from the outset. First and foremost, neutrino must be massive - thus the theory predicted neutrino mass some 25 years before its discovery. Second, the RH charged gauge boson $W_R$ has to me much heavier that its LH counterpart the SM $W_L$, the question is though how much heavier. It is interesting that in the case of 2 generations the theory would have provided a natural source of CP violation trough the RH charged currents, implying an upper limit $M_{W_R} \lesssim$ TeV, an exciting possibility the common origin of both P and CP violation. Instead, as we know, nature chose the existence the third generation - an example of a beautiful theory killed by ugly facts of nature?\\
   
  I go here through some essential features of the LR theory, but will not be exhaustive - for recent complementary brief reviews, see~\cite{Senjanovic:2023czt,Senjanovic:2025esv}).  For a more complete review, see \cite{Tello:2012qda}. The crucial aspect is the seesaw mechanism~\cite{Minkowski:1977sc,Mohapatra:1979ia,Yanagida:1979as,Glashow:1979nm,GellMann:1980vs}, with the RH neutrino mass matrix being proportional to the scale of LR symmetry breaking, i.e. the mass of the RH gauge boson, $M_N \propto M_{W_R}$, which in turns provides a small neutrino mass
\begin{equation} \label{eqSeesaw}
  M_\nu = - M_D^T \frac{1}{M_N} M_D,
\end{equation}
 where $M_D$ is the neutrino Dirac mass matrix. 
   Neutrino is light, because parity is strongly broken in charged weak interactions, and in the limit of the infinite $W_R$ mass
 one recovers massless neutrinos of the SM. This tells us why the fathers of the SM got the wrong prediction of massless neutrinos, in assuming 
   that parity is maximally broken as the experiment was suggesting - and maximally broken for all seasons.\\
   
   Moreover, neutrino is Majorana particle, leading to Lepton Number Violation (LNV), both at low and high energies exemplified by 
   \begin{itemize}
 
 \item Neutrinoless double beta decay,
 \item LNV at hadron colliders (KS process).
 
 \end{itemize}
   The former process, suggested soon after Majorana's great work, is by now a text-book example of LNV at low energies. Besides the usual neutrino contribution through the $W_L$ exchange, the theory predicts analogous contribution of RH neutrino N through the $W_R$ exchange, and this could make all the difference in the world. Imagine that you observe this process, and you measure the polarisation of the outgoing electrons  - and they come out RH. This would immediately tell you that the process is not due to neutrino exchange, but from the latter contribution of the RH sector of the theory. It is a straightforward exercise to see that then $W_R$ must be accessible at the next hadron collider, if not already at the LHC, since one gets $M_{W_R} \lesssim 20$ TeV~\cite{Nemevsek:2011aa}. \\
   
   This is a generic feature of any BSM theory, as can be estimate that from an effective operator for this process
  \begin{equation}\label{2betaeff}
  {\cal H}_{eff} = \frac {1}{\Lambda^5}\, \bar u\, \bar u\, \bar e\, \bar e\, d\, d \,.
   \end{equation}
The lack of observation of neutrinoless double beta decay $\tau_{0 \nu 2 \beta} \gtrsim 10^{25}\, yr$ can be shown to imply a lower limit 
   $\Lambda^5 \gtrsim 10^{18} GeV^5$, or  $\Lambda\gtrsim  4 TeV$. Of course, I simplified the analysis by assuming a single scale responsible for the process - in realistic situations there may be more scales as in the case of the LR theory, $m_N$ and $M_{W_R} $.
   I cannot overemphasise the importance of this result and the implication for our search for the BSM theory: if electron in neutrinoless double beta decay were to come RH, new physics would lie at energies on the order of 1-10 TeV. This is by far the most sensitive process to new scales. For example, the physics behind the Lepton Flavour Violation (LNV) could easily lie at energies of hundreds of TeV and produce observable results. LNV is our best bet for new physics reachable at today's or tomorrow's energies.  \\
   
   The main process exemplifying LNV at hadron colliders, present in basically all theories with Majorana neutrinos, is the so-called Keung-Senjanovi\'c (KS)~\cite{Keung:1983uu} process mentioned above, consisting of two same-sign charged leptons and two jets in the final state. This is complementary to the low energy neutrinoless double beta decay, and in this theory these processes are deeply connected. There is more to it, though. When RH neutrinos are produced, they must decay equally into charged leptons and anti leptons, allowing for the direct test of 
   their Majorana nature. So, the KS process is the high energy analogue of the neutrinoless double beta decay, with a great advantage of probing the physics behind the LNV. And in case of the outgoing electrons having the RH polarization in neutrinoless double beta decay, there is a deep connection with the KS process~\cite{Tello:2010am}.\\
   
      With the advent of the LHC, the theory has been studied in depth, and what emerged is quite surprising: LR model is completely self-contained predictive theory, just like the SM model itself. To be clear, it cannot predict its scale in the absence of new low energy phenomena, but if the scale of LR breaking were to be reachable at the LHC, there would be a number of computable physical processes. In what follows, we summarise briefly the main features of these remarkable predictions, both in the leptonic and quark sectors. \\
   
    \subsection{Leptonic sector: untangling seesaw mechanism}  \label{unsessaw}  
    
    The seesaw mechanism is admittedly beautiful in its message: neutrino is light due to its Majorana nature that sets it apart from its buddy the electron - its mass being inversely proportional to the mass of its RH counterpart, the heavy state N. But that {\it per se} is not sufficient to claim a theory of neutrino mass since we cannot predict neutrino mass itself. But then, one does not predict electron mass either - rather, one relates it a physical process, the Higgs decay into electron and positron, allowing us, as we said, to probe the origin of electron mass. The LRSM is in this sense completely analogous to the SM itself, it does for neutrino mass what the SM does for the electron one - it predicts $M_D$ from the knowledge of $m_\nu$ and $M_N$~\cite{Nemevsek:2012iq,Senjanovic:2016vxw,Senjanovic:2018xtu,Senjanovic:2019moe}, 
 and thus the mixing between light and heavy neutrinos. For this to happen, the LR symmetry is essential in restricting the form of $M_D$ - in general there is a huge ambiguity. The knowledge of $M_D$ is equivalent to the knowledge of Dirac Yukawa couplings, which, just as the Yukawa couplings of charged fermions in the SM, in turn determine physical decay rates. In particular, in case $M_N \gtrsim M_W $, one gets 
   \begin{equation}\label{NtoWe}
\Gamma (N \to W^+ \ell) \propto m_N \left(\frac{m_\nu m_N}{M_W^2}\right) \,,
\end{equation}
 where for simplicity I illustrated it for a single generation. All one needs to know are the Majorana masses of the LH and RH neutrinos, in analogy with the charged fermion mass in the SM. In the three generations case, the formula gets generalised with mass matrices, but that is not essential - what matters is that the former is being measured at low energies, and the latter at hadron colliders through the KS process.  In short,~\eqref{NtoWe} shows is the LRSM analogy of the celebrated SM result in~\eqref{higgstoffbar}.\\

   \subsection{Quark sector: determining RH charged current}  \label{wrcurrent}  
   
   The LRSM posed a great challenge from the very beginning. Although the LR symmetry is almost maximally broken in the low energy weak interaction, the theory should keep the memory of the original symmetry due to the fact that its breaking is of  spontaneous origin.  In other words, the RH quark current should in principle be related to the LH one, however, it took some 40 years to come up with an analytic formula. I will not dwell on its historical development, just summarise the final solution to the problem.
   The right-handed quark mixing matrix $V_R$ has a simple approximate  form~\cite{Senjanovic:2014pva} as a function of the usual left-handed CKM matrix $V_L$
\begin{equation}
\label{eq:VR}
(V_R)_{ij} \simeq (V_L)_{ij} - i  \epsilon  \frac{(V_L)_{ik} ( V_L^{\dagger}m_uV_L)_{kj} }{m_{d_k}+m_{d_j}}  
+O(\epsilon^2) 
\end{equation}
where $\epsilon $ is a small unknown expansion parameter. It can be then easily shown that the left and right mixing angles are almost the same, and right-handed phases depend only on $V_L$ and  
$\epsilon $. An interested reader should consult~\cite{Senjanovic:2015yea} 
 for more details regarding the right-handed mixing angles and phases.\\
 
 All in all, the LRSM is a highly predictive theory, with a plethora of physical process determined by the eventual knowledge of LH and RH neutrino mass matrices. Both CMS and ATLAS experiments are searching for the physics of the RH charged gauge boson through the direct hadron decays and the KS process,  providing a rough limit 
 $M_{W_R} \gtrsim 4$ TeV.  A nice summary of present day and future situation regarding the high energy physics of the LRSM can be found in~\cite{Nemevsek:2018bbt,Nemevsek:2023hwx}, while a careful study of precise predictions can be found in ~\cite{Kriewald:2024cgr}.\\

 \subsection{LRSM and cosmology: DM and leptogenesis}  \label{lrcosmo}  
 
  If we take LRSM as a serious candidate for the BSM theory, it is fair to ask if the theory can provide a natural DM candidate, and, in view of the Majorana nature of RH neutrinos, account for  leptogenesis. The answer to both question is yes, but it comes with a price. First, the lightest neutrino could be easily the DM, but then all three of them would have to be quite light, and their  direct observation through the KS process would be gone. Moreover, it seems that $W_R$ would have to be too heavy to be accessible at the LHC. - except, for a possibly small window around $M_{W_R} \simeq 5$ TeV~\cite{Nemevsek:2012cd}. 
  The leptogenesis, on the contrary requires the RH neutrinos to be quite heavy, but with $W_R$  necessarily out of the LHC reach.\\

 \paragraph{Dark matter}
 
 It takes a little thought ~\cite{Nemevsek:2012cd} to see that the only candidate for DM in the minimal LR symmetric theory is the lightest RH neutrino~\cite{Bezrukov:2009th,Nemevsek:2012cd} in the form of warm dark matter. This is the Dodelson-Widrow scenario~\cite{Dodelson:1993je} of sterile neutrino as the DM applied to the LRSM, with the following results
\begin{align}\label{flavorspectra}
	\mathbf{V}^R_\ell &\approx \begin{pmatrix}
		0 & 0 & 1
		\\
		0 & 1 & 0 
		\\
		1 & 0 & 0 
	 \end{pmatrix}, \quad \quad
	 \begin{array}{rl}
	 m_{N_1} & \sim \text{ keV},
	 \\
	 m_{N_2} & \approx m_\pi + m_\mu,
	 \\
	 m_{N_3} & \approx m_\pi + m_e\,,
	 \end{array}
\end{align}
 where $\mathbf{V}^R_\ell$ is the RH leptonic mixing matrix, and $m_{N_i}$ are masses of the three RH neutrinos $N_i$.\\
 
 \paragraph{Leptogenesis }
 
 It is well known that the RH neutrino Majorana  picture leads naturally to leptogenesis~\cite{Fukugita:1986hr}, and it goes through almost in the same manner - except that one must ensure that the  gauge interactions  of RH neutrinos with $W_R$ do not keep them in thermal equilibrium for too long. This leads to a lower limit on $W_R$ mass $M_{W_R} \gtrsim 30$ TeV~\cite{Frere:2008ct}.\\
 
 \subsection{And if neutrino was not Majorana?}  \label{lrdirac}  
 
  In its original version,  the LR symmetric theory actually employed new Higgs scalars  as $SU(2)$ doublets that led to the Dirac neutrino picture. The question was then what would made neutrino so much lighter that its weak partner, the electron, and the issue was studied at length 
  in ~\cite{Branco:1978bz}. This picture was abandoned with the advent of the seesaw mechanism in the version of the theory where the doublets were traded for triplets, which, as I have argued here, lead to a predictive theory of neutrino mass, with its smallness naturally accounted by the near maximal breaking of LR symmetry. However, in all fairness, for all that we know, the original picture could still be right, and the situation may be more interesting than imagined. Namely, in this case, in analogy with the quark sector, one can predict the RH leptonic mixing matrix~\cite{dartagnan} - impossible in the Majorana picture due to a completely different origin of LH and RH neutrino masses. This has impact even on the issue of strong CP violation~\cite{dartagnan}, interestingly enough. Since it is hard to probe the Dirac nature of neutrinos - for now one just keeps lowering the limit on neutrino Majorana mass - one may wonder if there is a direct test of this version of the theory at hadronic colliders. The answer is positive: in the Dirac case, the Higgs structure of the theory makes the mass ratio of new neutral and charged gauge  bosons smaller by a factor of $\sqrt 2$, compared to the Majorana case. I am not advocating here the Dirac scenario - I for one can bet on the Majorana version and observable lepton number violation - but it should be kept in mind. It is worth stressing that, if the symmetry is enlarged to include the Pati-Salam quark-lepton symmetry, the theory possesses magnetic monopoles~\cite{tHooft:1975psz} and it offers potentially observable gravitational waves, see~\cite{Senjanovic:2025enc} for a recent analysis. \\
  
   \subsection{LR symmetry and strong CP violation}  \label{lrtheta}  
   
   As I have stressed above, in the SM strong CP violation parameter $\bar \theta$ is completely independent from the rest of the theory and its smallness does not affect any of the parameters of the theory. The situation is very different in the LRSM, the reason being that the strong CP violation is akin to strong P violation, and affects the very LR symmetry used to define the theory. One could imagine that the original P symmetry suffices to control the amount of CP breaking~\cite{Mohapatra:1978fy,Beg:1978mt}, but it is not sufficient - the smallness of  $\bar \theta$ requires the smallness of the spontaneous part of the weak CP violation ~\cite{Maiezza:2014ala}, which then implies the smallness of the leptonic  CP violation~\cite{Kuchimanchi:2014ota}. This is turn puts an upper limit of masses of RH neutrinos~\cite{Senjanovic:2020int}, a welcome restriction since it helps control the smallness of lepton flavor violation for low scale of LR symmetry breaking. An interesting variation of this takes place in the version of the theory based on Dirac neutrinos~\cite{dartagnan}.\\

 \subsection{LR symmetry and domain wall problem}  \label{lrdw}  

   Before closing this section, I should stress that here left-right symmetry is not imagined to be an exact symmetry of nature - all that is assumed that its spontaneous breakdown dominates over the explicit ones. I did not wish to convince the reader of this principle, but only to demonstrate how it leads to a self-contained, predictive theory of neutrino mass. The reason that I, for one, still follow it, is that it predicted neutrino mass long before experiment confirmed it, and even more important, led to the seesaw mechanism behind its smallness. Of course, the explicit breaking may be there and may even be welcome, since even if it was minuscule, suppressed by the Planck scale, it could serve to rid us~\cite{Rai:1992xw} of the infamous domain wall problem~\cite{Zeldovich:1974uw}, associated with spontaneous breaking of discrete symmetries. This is important, since in the absence of inflation - and the minimal LRSM does not inflate - another way out of the domain wall problem~\cite{Dvali:1995cc}, the symmetry non-restoration at high temperature~\cite{Weinberg:1974hy,Mohapatra:1979vr,Mohapatra:1979bt}, unfortunately does not work in the minimal theory~\cite{Dvali:1996zr}. \\

  \section{Grand unification}  \label{gut}  
 
 Grand unification arose in the '70s as an attempt to unify all relevant particles forces (due to its weakness, gravity is irrelevant in this sense, at least as long as one stays sufficiently below the Planck scale) and to account for the quantisation of charge. This led from the outset to two remarkable predictions
 
 \begin{itemize}
 
 \item Magnetic monopoles. 
  
 It is worthwhile recalling here Dirac's classic prediction~\cite{Dirac:1948um}: the existence of a single magnetic monopole requires charge quantisation, a nature's miracle. Equally exciting, charge quantisation, automatically present in grand unified theories based on a simple single gauge group, necessarily implies existence of magnetic monopoles, as extended objects~\cite{tHooft:1974kcl,Polyakov:1974ek}. 
 
 \item Proton decay.
 
  Just as the W boson changes an electron into a neutrino since they live together in $SU(2)$, new superheavy gauge bosons of grand unified theories, change quarks into leptons and antiquarks since they all live together in a GUT gauge group - implying necessarily the decay of the proton.
  
     These predictions have become a kind of Holy Grail of modern particle physics, but alas without success. Proton is still stable for what we know, and monopoles have never been observed. Proton lifetime depends on a theory in question, and no model has been yet rejected based on proton longevity - more about it below. On the other hand, the existence of monopoles depends on the unknown cosmology at astronomical temperatures on the order of grand unification scale and as such cannot be used to constrain our theories. 
 
 \end{itemize}

     \subsection{Instructive failure: original $SU(5)$ theory}  \label{ggmodel}

 The minimal such theory, the original $SU(5)$ model of Georgi and Glashow,~\cite{Georgi:1974sy}. is an impressively predictive theory, actually too much for its own good. It predicts beautiful fermion mass relations $m_e = m_d$, and in turn all the proton branching ratios in terms of the same mixings that determine the weak interaction quark and lepton flavour transitions. This is seen manifestly in the interactions of the new heavy gauge bosons~\cite{Mohapatra:1979yj}
  \begin{equation}\label{xyint}
  \begin{split}
 {\cal L}_{X,Y} = X^\mu (\overline {u^c} \gamma_\mu u + \bar e \gamma_\mu d^c + \bar d \gamma_\mu e^c) +\\
  Y^\mu (\overline {u^c} \textcolor{red}{V_{CKM}} \gamma_\mu d + \bar \nu \textcolor{red}{V_{PMNS}} \gamma_\mu d^c + \bar u \gamma_\mu e^c )\,.
  \end{split}
 \end{equation}
 Remarkably, just as the W boson interaction, the new ones depend only on quark mixing matrix $V_{CKM}$ and lepton mixing matrix $V_{PMNS}$. Of course, this beautiful prediction is a product of the above wrong relation $m_e = m_d$, and as such is also necessarily wrong. I give here only to illustrate how a good theory should look like - the task of finding it turned out basically impossible. \\

 There is more trouble. The gauge couplings fail to unify and to make it worse, just as the SM, the theory predicted vanishing of neutrino mass.    This was quite a blow to the unification program but it paved a way for endless model building. In spite of my dislike for such a game, I can be blamed too for a model with Borut Bajc which in the context of $SU(5)$  cured the problem through the addition of $24_F $ fermion representation, which in turn predicts a light weak triplet fermion state~\cite{Bajc:2006ia,Bajc:2007zf,Arhrib:2009mz}.  There is a certain simplicity in the model, but still, it is an {\it a posteriori} model building, not a fundamental theory with predictions emerging from its structure. Another simple model worth mentioning~\cite{Dorsner:2005fq} is based on the addition of a symmetric Higgs representation $15_H $ instead of $24_F $, incorporating the so-called type II seesaw mechanism. 
     I for one would prefer a theory based on fundamental principles {\it a la} Feynman. \\

       \subsection{ $SU(5)$ theory only effective?}  \label{effectivesu5}

    Let us step back. The problems clearly stem from the simplicity of renormalisable d=4 structure - what about then accepting higher dimensional operators, and only higher dimensional operators, in the original theory? Of course, even this is a kind of betrayal of the original idea of the perturbative theory with structural predictions, but it has an appeal in its generality and simplicity. However, over the years,  it was proclaimed dead on the basis of the claimed too low unification scale, which was supposed to lead to premature death of the proton. What was premature, though, was only the death proclamation and decades of its necrologies - the theory is perfectly alive and well, albeit living dangerously~\cite{Senjanovic:2024uzn}. An improvement of proton lifetime by two orders of magnitude would definitely rule it out, and even one of order of magnitude, feasible by the next generation of proton decay experiments, would push the theory into a potion of parameter space,
    and imply the colour octet and weak triplet scalars from the adjoint Higgs representation to lie much below the unification scale. \\

     \subsection{ $SO(10)$ theory}  \label{so10}  

 In $SU(5)$ theory fermions are not unified but mistreated in 5 and 10-dimensional representation. This is cured in a theory based on $SO(10)$ gauge group~\cite{Fritzsch:1974nn,Georgi:1974my}, where the SM fermions, together with a necessary RH neutrino, reside in a $16_F$ spinorial representation.  The crucial point is that the $SO(10)$ theory contains the LR symmetric electroweak part, just as $SU(5)$ embeddies the SM gauge group.  It also contains the Pati-Salam quark-lepton unification based on $SU(4)$ symmetry,  an important property but not of our interest here. What is crucial for our purpose is that the LR symmetric is automatic, and it is a finite gauge transformation of $SO(10)$.  The problem is, just as in the minimal $SU(5)$, one gets wrong mass relations $m_e = m_d$ - yet another beautiful theory killed by ugly facts of nature?  The answer is no if one believes in the necessity of higher dimensional operators as in the minimal $SU(5)$ theory, or adding a large Higgs representation $126_H$, which can both lead to tree level seesaw mechanism and correct mass relations~\cite{Babu:1992ia}. \\
 
     Since $126_H$ proliferates the scalar spectrum of the theory and prevents making clear predictions, it is worthwhile to investigate a possibility where the role of $126_H$ is played by a spinorial $16_H$, and appeal to higher dimensional operators. Then, in analogy with the $SU(5)$, one arrives at realistic theory, with the mass of the RH neutrino being generated both by a d=5 operator and a two-loop effect~\cite{Witten:1979nr} (for a review of seesaw mechanism, with emphasis on unified theories, see~\cite{Senjanovic:2011zz}). The situation is subtle, however, since the gauge symmetry kills d= 5 operators in the minimal version of the theory, where the role of $24_H$ field of $SU(5)$, needed for GUT symmetry breaking, is played by the anti-symmetric $45_H$ dimensional representation. The smallness of $d=6$ operators, doubly suppressed by
 $M_X/\Lambda$ poses the challenge and unification of gauge couplings, and the theory works nontrivially.  Still, it does not need the flavour rotation of proton decay and a practitioner of grand unification, who is unhappy with such fine-tunings, will be relieved to know that theory can still work - and the way it does, is to predict light colour octet and weak triplet scalar particles at low energies, potentially accessible even at the LHC~\cite{Preda:2022izo}.\\
 
    I have shied away here from renormalisable grand unified models which require large Higgs representation, butt they areviable alternatives in their own right for their Yukawa sectors are less complex and often more predictive. An interested reader can consult~\cite{Senjanovic:2011zz} for a review of such models.  I should add that in recent years there has been an important progress in the minimal renormalisable $SO(10)$ theory, made  by the authors~\cite{Jarkovska:2023zwv}., who claim that  the theory based on the $45_H$ and $126_H$ representations is  ruled out- a remarkable result, if true. If still viable, however, it could lead to interesting low energy processes and new light states, as argued in~\cite{Preda:2025afo} In any case, there is always a possibility of $45_H$ being substituted with a symmetric $54_H$ representation, with a different pattern of symmetry breaking and different cosmological implications.\\
 
  \subsection{Grand unification and supersymmetry}  \label{susygut}  
   
     As I said in the Prologue, low energy supersymmetry found its natural place in grand unification, and for a long time even took over ordinary, original, version of the idea. First, it allowed for just the tree level fine-tuning of the Higgs mass parameter once for all, and second it prophetically predicted correct value of the weak mixing angle.  The former point at the end is not so important, for whether one likes or not, in the minimal theories one still must fine tune. Second, in the context of $SO(10)$ GUT unification of couplings can proceed equally naturally through the existence of the intermediate mass scale responsible for the generation of neutrino mass. But even in $SU(5)$ theory nothing really forces the supersymmetry breaking scale - except for a desire to be low - since unification of gauge couplings allows it to be huge, close to the GUT scale itself~\cite{Senjanovic:2023jvv}. \\

      \subsection{Forget GUT, just effective?}  \label{effectivesu5}  
      
      As I tried to show, the moral from all this is simply that grand unification has not yet come up with a theory that could make Feynman happy. Its central aspects, the nucleon decay and the existence of magnetic monopoles still lack predictivity. When it comes to monopoles, it all depends at the unknown picture of the universe at enormous temperatures on the order of the GUT scale. Or better, universe may have never been that hot, and even it had been, there may not have been thermal equilibrium needed to establish the phase transition to the unbroken symmetry phase - and it is precisely such a phase transition that leads to the creation of monopoles through the so-called Kibble mechanism~\cite{Kibble:1976sj}. Regarding nucleon decay, besides the original Georgi-Glashow theory ruled out by experiment, no other theory managed to predict its branching ratios, with the same structural clarity (for recent attempts in this direction, see e.g.~\cite{Babu:2016bmy,Dorsner:2024jiy}). \\
      
      There is one clear cut feature of grand unification, the enormity of its scale. What happens if one gives up - at least for time being - probing specific models? Could one devise general ways to test some generic features of nucleon decay from the assumption of its physics stemming from scales much larger that the weak scale? In other words, if one were to go effective all the way, would there be nucleon decay features that any UV completion would have to satisfy? This is answered positively by Weinberg~\cite{Weinberg:1979sa} and Wilczek and Zee~\cite{Wilczek:1979hc} in their beautiful analysis, that leads to a number of well defined sum rules.\\
      
        In the large scale picture $\Lambda \gg M_W$, it turns out that there are only four leading d= 6 operators (see e.g.~\cite{Senjanovic:2009kr} for a review)
\begin{equation}\label{Beffective}
\begin{split}
\mathcal{O}_1 &= (u_R  d_R) (q_L \ell_L),  \,\,\,\,\,\,\,\,\,  \mathcal{O}_2  =  (q_L  q_L)  (u_R  e_R)\\
\mathcal{O}_3 &= (q_L   q_L) (q_L \ell_L),  \,\,\,\,\,\,\,\,\,\,\,  \mathcal{O}_4 =  (u_R  d_R)  (u_R  e_R)\,.
\end{split}
\end{equation}
It is easy then to derive a set of immediate isospin relations between proton and neutron decays into both charged  leptons and neutrinos, and neutral and charged pions, see e.g~\cite{Senjanovic:2009kr} .\\
        
        Also, one can show that there is an accidental  $U(1)_{B-L}$ global symmetry (difference of baryon and lepton numbers), which says that, among other things, 2-body neutron decay final states cannot contain charged leptons, only their anti particles.  It gets even more interesting in case of neutron decay into kaons since the operators in \eqref{Beffective} imply anti strange quark in the final state, or $K^+$ - but then the accompanying lepton would be negatively charged, in contradiction with B-L symmetry. In other words, if baryon number violation stems from high energy physics, as suggested by the idea of grand unification, neutron two body decays cannot involve charged kaons
\begin{equation}\label{ndecayintoK}
n \not\to K^+ \ell,  \,\,\,\,\,\,\,\,\, n \not\to K^- \bar \ell
\end{equation}
         This is  a remarkable result: observing such decays would invalidate the whole idea of conventional grand unification. Of course, one could circumvent this by having some scalar particles that mediate proton decay lie at low energies, by fine-tuning their Yukawa couplings to be sufficiently small. In particular, this has been advocated by Dvali in the case of the colour triplet of the SM Higgs doublet~\cite{Dvali:1992hc}. \\

      \section{ Summary and outlook for the future}  \label{Outlook}  
      
      If the reader managed to arrive at this point, they hopefully share my conviction that natural phenomena are studied through well defined predictive theories based on fundamental principles. This is what history teaches us, this is what research truly means and this is what makes it so special. Of course, it is experiment that decides at the end, but the process of selecting fundamental theories of nature and then testing them, is all what science is about. The trouble, of course, is to know what theory to work on, and moreover how to actually verify it - but what does not help is endless model building, where one adds in an {\it ad hoc} manner particles and their interaction to existing theories. To make it worse, one often claims minimality by counting such additions, in spite of no guiding principle behind them - as if numbers could decide what is fundamental.  \\
      
      
      And there are such theories today, just as the Standard Model was around in the sixties when instead everybody was pursuing shallow ideas of bootstrap and such. I gave examples, based on my personal bias, of the Left-Right symmetric theory of electroweak interactions that turned into a predictive theory of neutrino mass, and both minimal $SU(5)$ and $SO(10)$ grand unified theories augmented by higher dimensional operators. The LRSM should be of immediate interest to experimentalists since it predicts a plethora of high and low energies processes. I have focused here on the KS process, since it serves both as a paradigm of LNV at hadron colliders, and a way of probing directly the Majorana nature of RH neutrinos - which in turn allows for the direct probe of the seesaw origin of neutrino mass. Moreover, there is a direct and deep connection between the KS process and low energy neutrinoless double beta decay, which makes all of this even more exciting.  And most important, something I cannot overemphasise: if we discover neutrinoless double beta decay and if the outgoing electrons were to turn our RH, then this process would provide a convincing argument in favour of LRSM, and moreover, its scale would have to lie at the next hadron collider, if not already at the LHC. \\
      
      I should stress here that it is easy to rule the LR theory out, at least in principle - all it takes is find a new neutral gauge $Z_R$ boson at the LHC. It turns out that  $M_{Z_R} \simeq 1.7 M_{W_R}$, and the existing limit  $M_{W_R} \gtrsim 4$ TeV makes the neutral analogue too heavy to lie at the LHC. 
      
     In short, the LR symmetry, besides predicting neutrino mass long before its experimental confirmation, provides a deep organising principle. and it touches into the core of the SM physics, the violation of parity on which it was built. I wish Feynman were around, I would love to know whether he would find the LRSM pass his criteria of a good theory. \\

        From the pure theoretical point of view, grand unification is even more appealing since it leads to two fundamental predictions, the decay of the nucleon and the existence of magnetic monopoles, the latter arguably the most exciting and most beautiful prediction of all modern particle physics - after all, almost every child in the world gets fascinated by the non-existence of single poles of magnets. The trouble though is that the minimal such theory, the original model of Georgi and Glashow, an analog of the SM and the LRSM, fails miserably - it does not even unify, ironically. It suffers from the same weakness of the SM, the structural prediction of zero neutrino mass - so maybe it serves it right not to pass other experimental tests? Its natural counterpart, the minimal $SO(10)$ theory, is, on the other hand, a beautiful theory of neutrino mass, through the same seesaw mechanism as the LRSM, as long as one allows for higher dimensional operators or large Higgs representations. In the former case, it suggests the existence of new light scalars at today's energies, but just as the in the latter case, it stops short of predicting nucleon decay branching ratios, its primary duty. On the other hand, the fate of magnetic monopoles depends on the unchartered territory of the early universe, and almost nothing can be said about their number here on Earth. \\
        
        At the time I am writing this, no truly predictive grand unified theory emerged. After more than fifty years, we still keep building models, and I too can be blamed for that too.  As much as I love the idea of grand unification - it was a love on first sight and I have been in love ever since - I, for one, remain deeply unhappy with its failure to be more predictive, at least in its simple form.  Not all is lost, though - there is still a way of testing its prediction of huge unification scales by doing an effective analysis of nucleon decay interactions, as reviewed in ~\cite{Senjanovic:2009kr}. \\

%

 \subsection*{Acknowledgments}
  
   I am grateful to the organizers of the LHC days in Split 2024 international conference,  Hvar, October 2024.  I wish to thank to Damir Lelas, Nikola Godinovi\'c,  Alejandra Melfo,  and specially Vladimir Tello and Michael Zantedeschi for useful discussions and comments,  their encouragement to write this up and careful reading of the manuscript. 
    I acknowledge warm hospitality of the friendly staff of Briig hotel in Split, where most of this essay was written.

\end{document}